\documentclass[useAMS,usenatbib,usegraphicx]{mn2e}
\usepackage{times}
\title[Frequency and time profiles of noise storm bursts]
  {Frequency and time profiles of metric wave
  isolated type I solar noise storm bursts
  at high spectral and temporal resolution}
\author[Shanmugha Sundaram \& Subramanian]
 {G. A. Shanmugha Sundaram$^{1, 2}$ \thanks{Email : sga@physics.iisc.ernet.in}
 and K. R. Subramanian$^2$\\
 $^1$Joint Astronomy Programme, Department of Physics,
 Indian Institute of Science, Bangalore - 560012, KA, INDIA.\\
 $^2$Indian Institute of Astrophysics, Koramangala,
 Bangalore - 560034, KA, INDIA.}
\begin{document}
\date{Received 2005 February 2; Accepted yyyy mm dd}
\pagerange{\pageref{firstpage}--\pageref{lastpage}} \pubyear{2004}
\maketitle
\label{firstpage}
\begin{abstract}
Type I noise storms constitute a sizeable faction of the
active-Sun radio emission component. Observations of isolated 
instances of such bursts, in the swept-frequency-mode
at metric wavelengths, have remained sparse, with several unfilled regions in
the frequency coverage. Dynamic spectra of the burst radiation,
in the 30 - 130 MHz band, obtained from the recently commissioned
digital High Resolution Spectrograph ( HRS ) at the
Gauribidanur Radio Observatory, on account of the superior
frequency and time resolution, have unravelled in explicit
detail the temporal and spectral profiles of isolated bursts.
Apart from presenting details on their fundamental emission features,
the time and  frequency profile symmetry, with reference
to custom-specific Gaussian distributions, has been chosen as the
nodal criterion to statistically explain the state of the source regions in the
vicinity of magnetic reconnections, the latent
excitation agent that contributes to plasma wave energetics,
and the quenching phenomenon that causes damping of the burst emission.
\end{abstract}
\begin{keywords}
plasmas -- radiation mechanisms: non-thermal --
instrumentation: spectrographs -- methods: data analysis --
Sun: radio radiation
\end{keywords}
\section{Introduction}
Type I radio noise storms head the list of solar events discovered
at metric wavelengths~\citep{kbab95}.
Noise storms rage over a prolonged duration, and appear as intense,
narrow-band bursts, superposed on a low-intensity
broadband continuum, in the 30-400 MHz range.
Either component of the noise storm radiation has a
very high degree of ordinary-mode
circular polarization ($\sim 100\%$~), and is widely believed to be
generated by the plasma emission mechanism.
Ever since their discovery~\citep{hey46}, theirs has been among
the extensively studied of solar radio features.
Over the years, several notable theoretical and observational works
on this particular phenomena have been performed \citep[see][]{kms85,
bewen81, mel80, mer80, krg79, mes77, elg77, man76, ker73, kun65, wam50}.\\
\\
Features such as short duration, high polarization, narrow bandwidth,
a random distribution in frequency, and height associated with
type I noise storm bursts, have been
considered as evidence for the radiation process to be of plasma
origin, occurring
in the vicinity of the local plasma frequency or at its harmonic(s).
The bursts are caused due to excitation of the upper hybrid
waves by a trapped population of energetic nonthermal electrons,
bounded by the magnetic field lines
of the coronal plasma~\citep{bak01}.
The electrons that are in the vicinity get
accelerated along the magnetic field, and such an anisotropic beam
tends to become unstable.
The bursts occur when nonthermal electron densities approach
a threshold value,
and the phenomenon exhibits significant sporadicity.
The threshold density values show a downward trend with frequency,
and fluctuations
in the densities of either the super-thermal electron beam or the
ambient coronal plasma, determine the bandwidth
and duration of type I bursts.\\
\\
We present observational results on this rapidly changing time and
frequency component of the metric type I noise storm radiation,
along the lines of their dynamic spectral
characteristics and the implied plasma emission mechanism.
The section that follows is an account of the
HRS, and the flux calibration scheme adopted for the observed
dynamic spectrum of bursts.
Unique traits associated with the isolated bursts, such as
bandwidth of emission, duration, and flux distribution,
are interpreted
in terms of the coronal plasma environment in the associated
source region. A statistical study is performed on the distribution of
frequency and time profile shapes, with reference to
customized Gaussian profiles, in order to
explain the excitation and damping mechanisms of plasma waves,
and the agent responsible for shaping the observed profiles.
The last section explains the spectrum and profile shapes,
from discussions based on
the strikingly corroboratory nature of the isolated metric bursts with
spikes observed at relatively higher frequencies,
with generic attributes to elementary radiation processes in the coronal
source regions of this plasma emission phenomenon.
\section{Instrument details \& Observations}
\subsection{High Resolution Spectrograph and Antenna System}
The HRS includes one group
of antennas of the Gauribidanur Radioheliograph (GRH~\citep{grh})
and a spectrum analyzer.
An array of four log-periodic dipole antennas (LPDAs) with their
E-planes aligned mutually ( along the E-W direction ) and
stacked along the H-planes ( along the N-S direction ),
constitutes the basic element for signal interception.
The (E-W) beam is ${90^{o}}$, and this offers about
6 h of continuous spectral observations
of the Sun in the frequency band of 30-150 MHz. The beamwidth
along the (N-S) direction varies from
${6^{o}}$ at 150 MHz to ${30^{o}}$ at 30 MHz;
hence the Sun remains practically unresolved over the entire
operating frequency
range of the spectrograph.\\
\\
The amplified radio signals are fed to a computer-controlled,
commercial off-the-shelf
(COTS) spectrum analyzer, that accomplishes
the role of multi-frequency solar radio spectrum analysis.
The device has a logarithmic response, with a frequency
sampling rate of about
41 ms. The instantaneous bandwidth ( or frequency resolution of
the spectrograph )
- which for this particular investigation on type I bursts
works out to ${\approx{250}}$ kHz.
A GPIB/RS-232 at the communication port enables acquisition
of data by a desktop computer
in a 3-way frequency vs. time vs. signal amplitude
( arbitrary units ) format termed the \emph{dynamic spectrum},
for offline spectrum analysis.
\subsection{Dynamic Spectral data and Flux calibration}
The output of the HRS is of a data type,
that has the detected signal
in arbitrary units of signal power, acquired in the frequency range
30-130 MHz, across the
401 equally-spaced frequency channels, for a duration of 6 h.
This transit-mode
observation schedule for the Sun has been in vogue since 2002 June,
with each day's
run commencing at 03:30 UT and ending at 09:30 UT.
Dynamic spectra of Sun were obtained on days 15, 16, 17, 18, 28~\&~30
of July and 13, 17~\&~18 of August 2002, since type I burst activity was
found to be very intense on those days.\\
\\
The galactic background has been used in gain calibration of the
spectral channels. Appendix A describes the procedure adopted for
calibration in greater detail.\\
\\
A suitable algorithm was deviced to 'harvest' isolated cases of
type I noise storms bursts,
that appear as intensity-enhanced bursts in the frequency-time plane,
the criterion for classifying the desirable spectral event being
its enhanced intensity
level as against the background, its emission duration and bandwidth.
The broadband continuum component of the noise storm spectrum was
essentially subtracted as
background radiation from the dynamic spectral records, as were the
long time-scale
modulations; this enhanced the dynamic
range for detection of the burst component in noise storms.
\section{Characteristics of the isolated noise storm bursts}
\begin{figure}
\resizebox{\hsize}{!}{\includegraphics{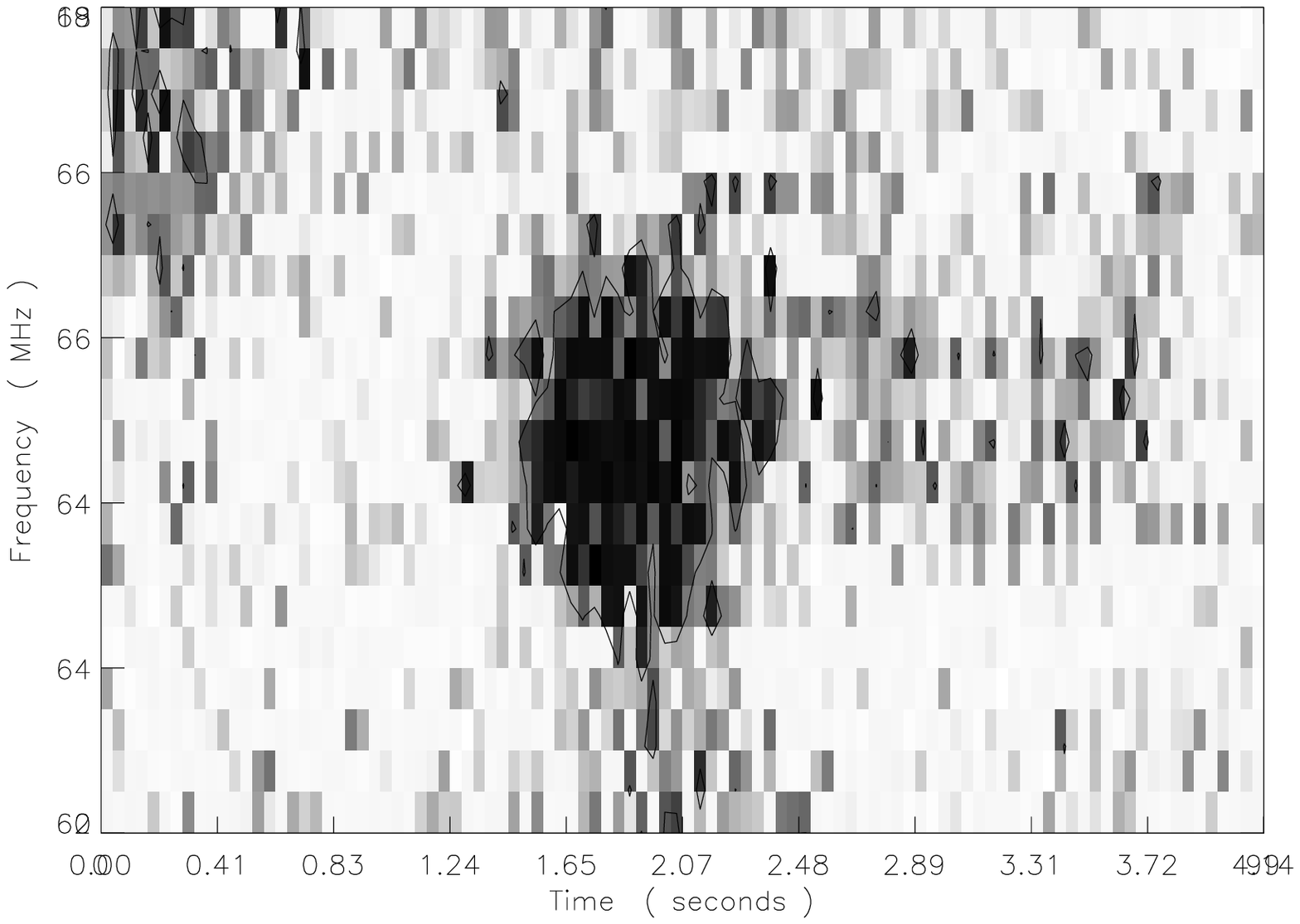}}
\resizebox{\hsize}{!}{\includegraphics{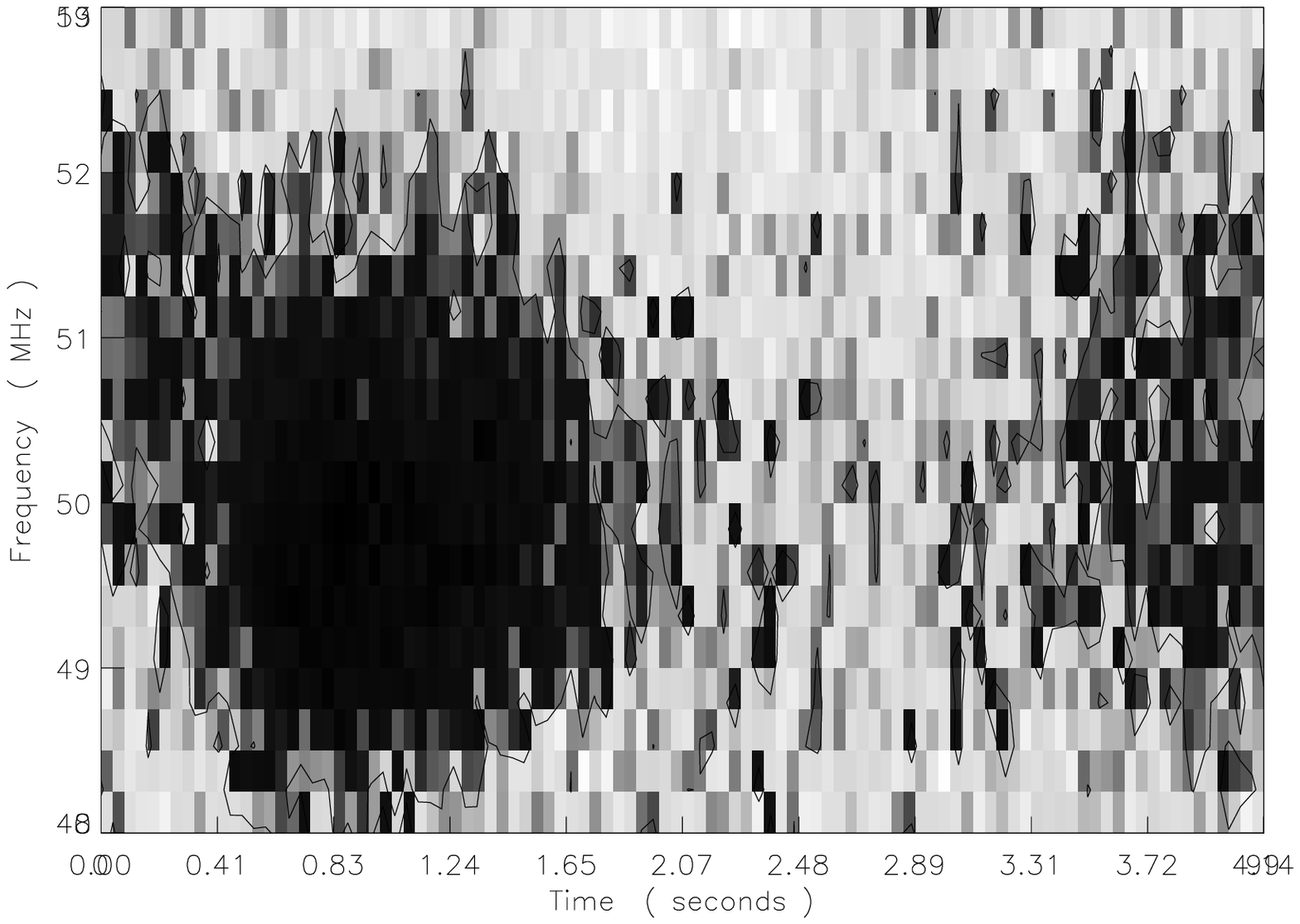}}
\caption{Dynamic spectral records of isolated Type I bursts
that occured on $2002 July 28^{th}$.}
\label{6fgh}
\end{figure}
In this section, the characteristics of isolated instances of type I
bursts, of the kind shown in Fig. \ref{6fgh}, such as their
distribution with central frequency, bandwidth,
duration ( lifetime ), and calibrated flux,
observed in the
dynamic spectral records for the few select days of 2002 July \&~
August, are reported. A collection of
39 individual instances of isolated bursts
was chosen for the study; they were well resolved in time and
frequency, and the compilation includes only bursts with profiles
that occured unambiguously in the expected typical bandwidth range.
The peaks were determined by looking for strong enough local
maxima above the noise level (${\ge 5 \sigma}$),
and a constant background
( representing the quiet-Sun and other slow-varying baseline
components ) was subtracted.
In the broad frequency range spanning 100 MHz, the study of frequency
and time profiles
of these rapidly evolving structures,
recorded at a high digitization rate, would be fundamental
to the analysis of profile shapes in terms of
the rise and decay phases, and hence provide
details on the source region for plasma emission.
\subsection{Distribution of radio flux of isolated type I bursts}
\begin{figure*}
[a]{\includegraphics[totalheight=2.39in]{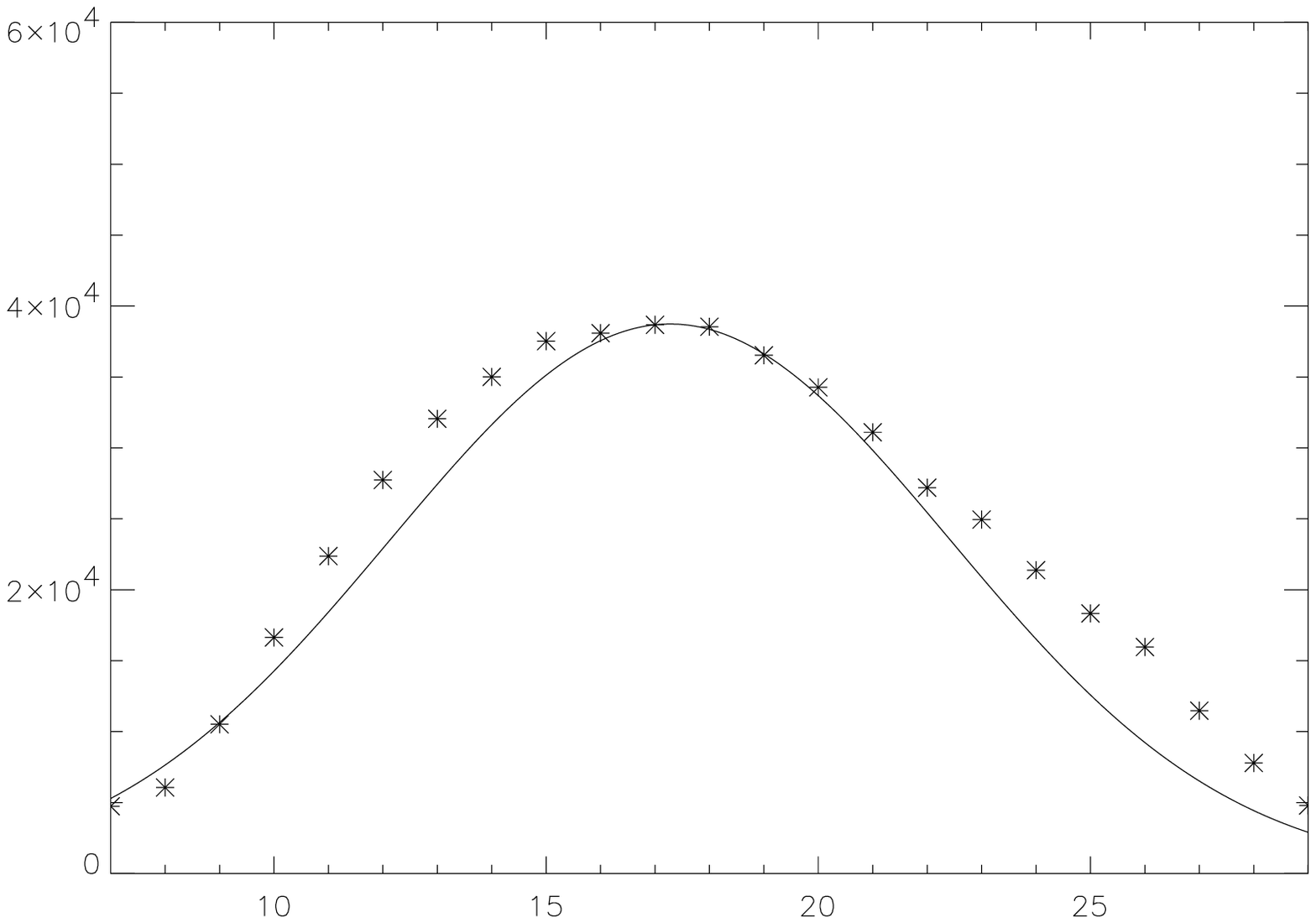}}
[b]{\includegraphics[totalheight=2.39in]{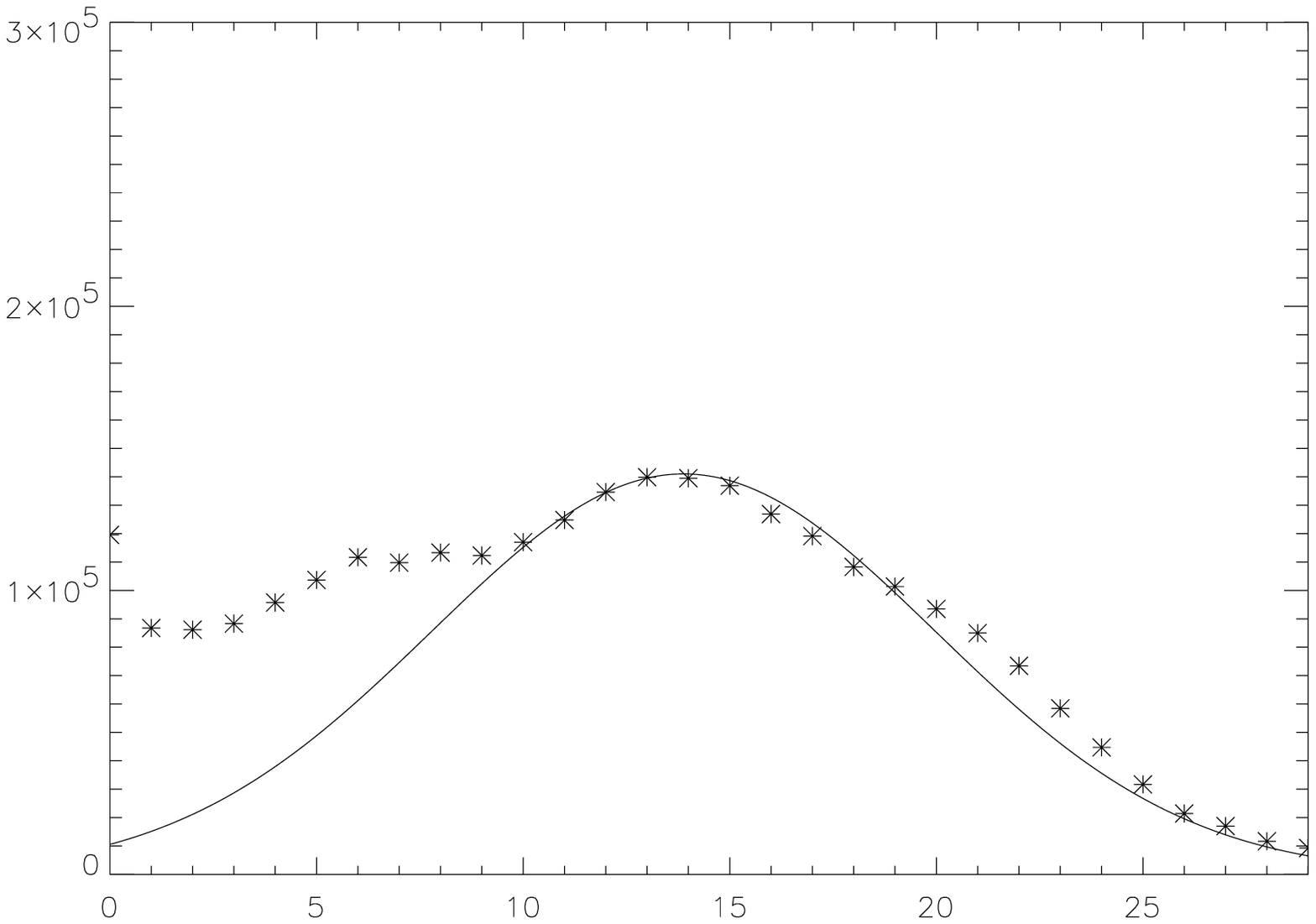}}
[c]{\includegraphics[totalheight=2.39in]{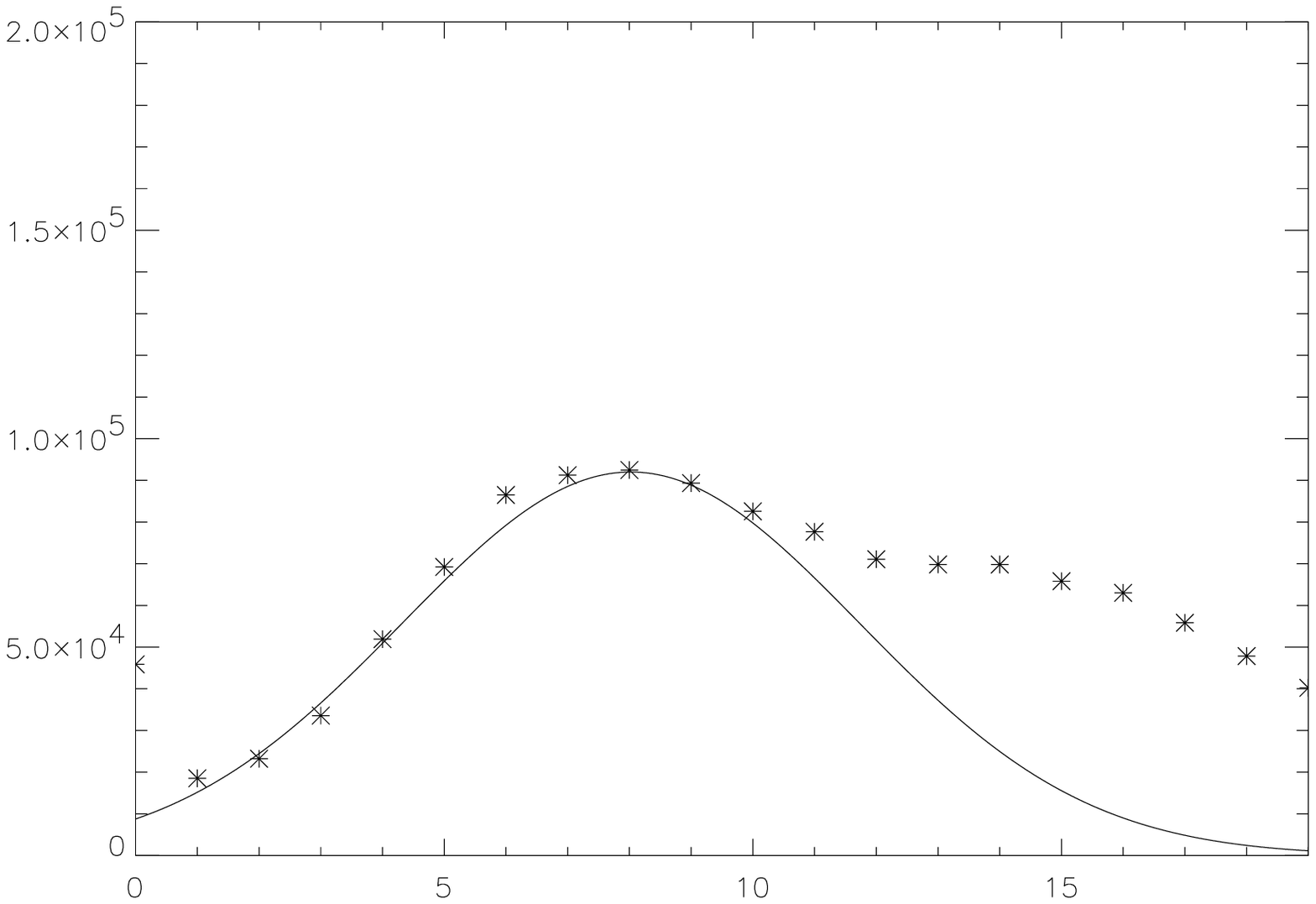}}
[d]{\includegraphics[totalheight=2.39in]{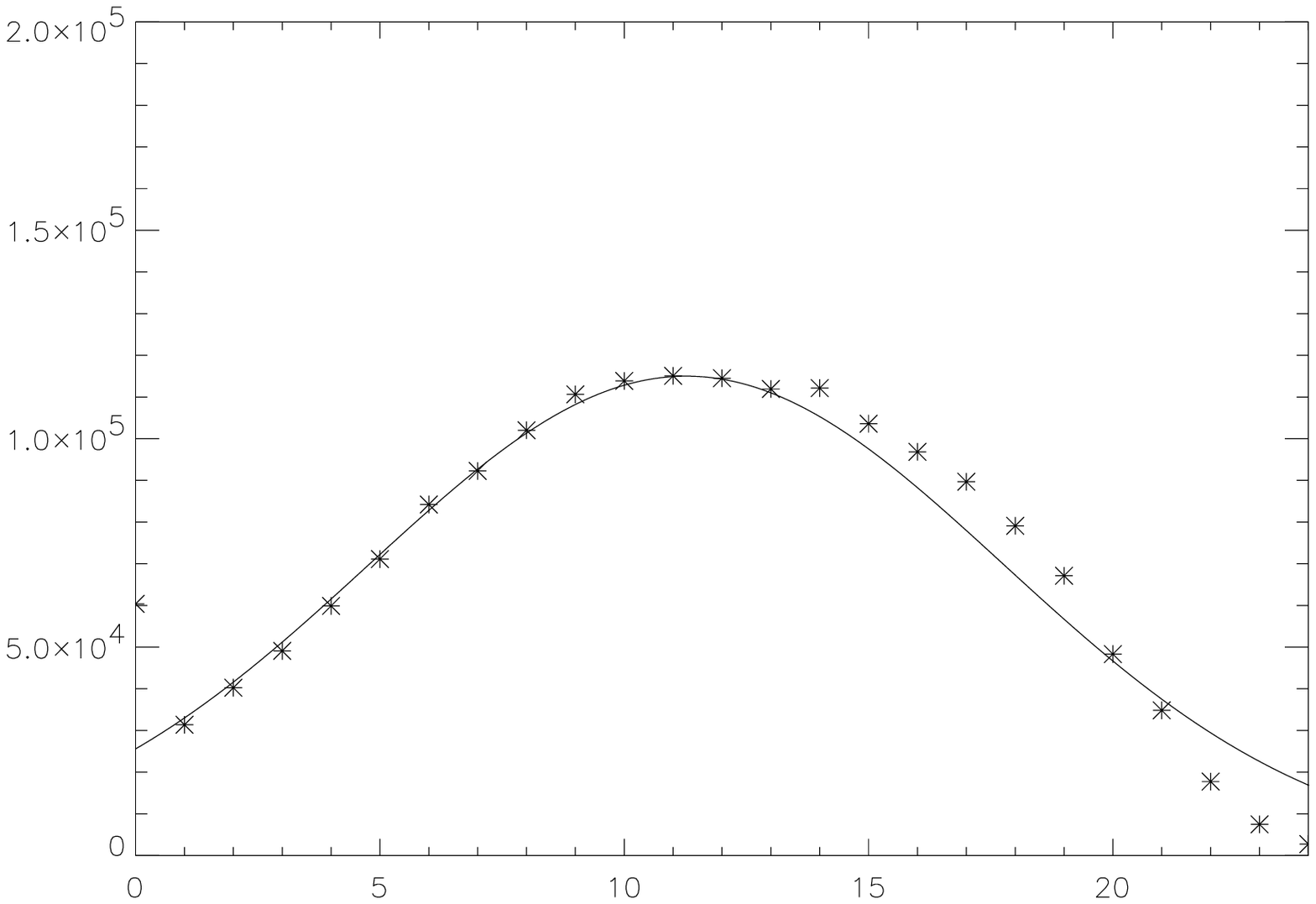}}
\caption{The 3 kinds of isolated type I noise storm bursts,
according to their time-averaged frequency profiles.
The abscissa is in terms of frequency-channels,
with adjacent channels being 250 KHz apart;
the ordinate axis is in uncalibrated flux units.
Sub-plots (a) \&~(d) indicate symmetry, while
(b) \&~(c) show trailing \& leading edge
assymetry, to the fitted gaussian distribution.\label{6fg7}}
\end{figure*}
The distribution of isolated type I noise storm bursts with their
calibrated radio flux, reveals a peak in the
distribution at $\mathrm{\sim 25}$~sfu.
The result concurs with a similar peak flux occurrence
amid 20-40 sfu
in the 70-130 MHz band~\citep{wld50}. There occurs an inherent uncertainity,
accociated with the dynamic range of the spectrograph,
in the detection and estimation
of low-intensity bursts' radio flux, during periods of intense noise storm
activity; this would leave the flux-poor bursts as being
unaccounted for, during such instances.
Yet, it has little impact on the detection of flux-rich bursts,
thereby implying that a
significant number of bursts during the period of noise storm
activity considered,
are of the low-intensity type.\\
\\
The narrowband bursts exhibit
frequency independent energy characteristics
at metric wavelengths;
about 75 \% of the observed bursts occur in the
42 - 85 MHz range.
This trend can be interpreted in terms of the
high degree of constancy in the density and temperature parameters,
along the path of the exciting disturbance in the active source region,
and in a direction parallel
to that of the solar surface ( as with the case of bursts
generated near the apex
region of evolving magnetic loops ).
\subsection{Frequency-profile analyses}
\subsubsection{Distribution of isolated bursts with
central frequency and bandwidth}
A distribution was noticed
in the isolated bursts as a function of
their bandwidth and fractional-bandwidth.
Of the 39 cases, more that half
(i.e. $\mathrm{54 \%}$) the
bursts have a bandwidth that is below 3 MHz;
69 \% below the 3.5 MHz, and
82 \% below 4 MHz. These values compare favourably with 2.8 MHz,
observed for the half-power bandwidth
in the 70-130 MHz range (\citet{wld50, elg77} and references therein ),
for the case where a Gaussian distribution is the most likely
frequency-profile fitting.
This fairly indicates that the majority of the isolated
type I bursts in this sample,
were emitted at half-power bandwidths less than
4 MHz, with the peak in the distribution being in the 2-2.5 MHz
restricted band.
Another notable feature was the near absence of
fundamental-harmonic
pairs among the burst events chosen for this study.
Since the bandwidth per channel
( $\mathrm{\approx 0.25 }$~MHz ) is considerably lower than the observed
bandwidths, significant
errors on this aspect can be discounted.\\
\\
The fractional-bandwidth ( FBW ) is defined as follows~\citep{tay95} :
\begin{eqnarray}
\Delta \nu_{fr}~=~2 \cdot (\frac {f_{H}-f_{L}} {f_{H}+f_{L}})~
\equiv~\frac {\Delta f} {f_{c}}
\end{eqnarray}
where $\mathrm{f_{H},f_{L},\&~f_{c}}$~are respectively the high, low and
central frequencies ( in MHz ) for each burst event, and
$\mathrm{\Delta_{f}}$ is its bandwidth.
The peak, in the distribution of burst events with FBW, for
$\mathrm{\sim~50~\%}$~of samples considered, occurs at an
instantaneous FBW of 4 \%, and this meets
the criterion, for the type I bursts considered, of narrow-band
emission characteristics.\\
\\
Analysis of dynamic spectral records on noise storm bursts indicates
an extremely low degree of frequency drift, and the absence of harmonics.
A quantitative study was performed,
in order to estimate the degree of deviation from symmetry, in cases where
Gaussian fittings were performed to the time-averaged frequency profiles.
A typical burst profile in frequency has a shape that favours the choice of a
Gaussian fitting, since (i) the spectrum of a burst is
symmetric in most cases,
at least within the frequency domain of the half maximim points
( though deviations from the gaussian curve are noticeable,
especially at the wings ) (ii) The spectral behaviour of the flux,
either side of the peak frequency, appears bell-shaped.
The Gaussian functions are of the standard form :
\begin{eqnarray}
g(x)~=~\frac{1}{\sigma \sqrt{2 \pi}}~exp \left[ - \frac{1}{2} \left(
\frac{x}{2 \sigma} \right)^{2} \right]
\label{6eq3}
\end{eqnarray}
where, $\sigma \sqrt{2 \pi}$~is the weighting factor in the distribution,
and the standard deviation $\sigma$ influences the gaussian
width, which in this case would be the half-power bandwidth
of the burst event, 'x' being the length scale of the distribution.\\
\\
Fig. \ref{6fg7} depicts the 3 kinds of profiles encountered in
the study viz., the symmetric ( sub-plots a \&~d ),
the ascending-limb assymetry ( sub-plots b) and the
descending-limb assymetry ( sub-plots c).
Asymmetry in the frequency profiles is found manifested
as an enhanced tail portion straggling the leading or trailing
limb of the gaussian profile fitted to the time-averaged data points.
Among the type I bursts detected, that conforms to one of the possible three
distribution patterns,
a majority ($\approx 54 \%$) of events follow a symmetric,
gaussian frequency
distribution, while ($\approx 26 \%$) of the lot belong to the ascending-limb
asymmetry, and ($\approx 21 \%$) to the descending-limb asymmetry.
The insenstivity of the bandwidth on frequency, and the occasional
observation of extremely small bandwidths, are reflected in the
near constant scatter of the fractional-bandwidth with the central
frequency of the bursts. This strongly indicates that the emission process
is of the narrowband type, and that the observed bandwidth of a burst is
its natural width due to the emission process and not a consequence
of inhomogeneities in the source region.
\subsection{Time-profile analysis}
\subsubsection{Distribution of lifetimes of isolated bursts}
\begin{figure}
\resizebox{\hsize}{!}{\includegraphics{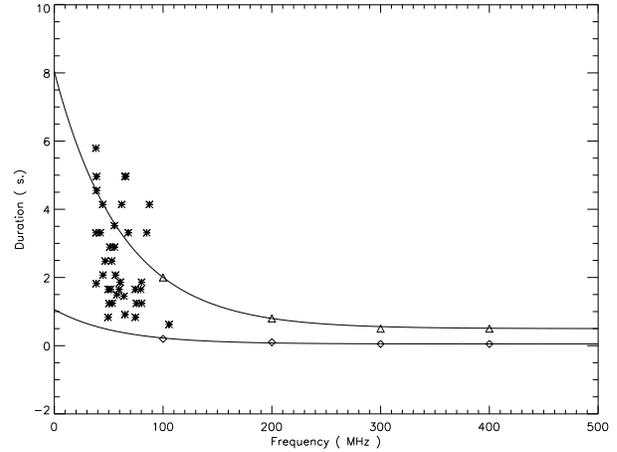}}
\caption{Duration of isolated type I noise storm bursts as a function
of frequency. The upper and lower curves are exponential-decay fittings
to observations as in~\citep{elg66}. Asterixes stand for the HRS
observations.}
\label{6fg5}
\end{figure}

Duration of a burst may be
given by the time interval between its very beginning ( which is usually
fairly well defined ) and its fall below the $\mathrm{3 \sigma}$
limit of the instrument, above the background level.
This would describe the total duration, homogeneously restricted
to a given sensitivity. Among the
bursts chosen with definite start and end times, the differing peak flux
levels, as well as subjective estimates of the background,
cause a considerable scatter of the result, which is referred to as the
"total duration" of isolated bursts.
The duration measures widely used in literature are : time interval
between the half peak flux levels ( half power duration ), time interval
between 1/e-peak flux levels ( natural duration ),
and time interval between
1/10-peak-flux-levels~(\citet{gab90} and references therein ).\\
\\
The natural duration is on an average close to  the physically meaningful
duration of the exciting agent arising as a consequence of the plasma
instability in the source region for the bursts.
The peak in the distribution of
isolated type I bursts with their lifetime occurs at 1.5 seconds,
and this agrees with values of 1.2 seconds~\citep{war69}, 1.5 seconds
(~\citep{wld50} in the 70-130 MHz band) and 1.01 seconds~\citep{sas69},
for the lifetime of bursts of type I noise storm radiation.\\
\\
The variation in mean duration of type I bursts
with their frequency of observation is depicted in Fig. \ref{6fg5}.
The upper and lower exponential-decay curves are fittings made to the
observations
of noise storm bursts in the 100-500 MHz frequency range
(~\citet{man76,elg66},~\& references therein );
they indicate the upper and lower limit on the duration
of the bursts. The type I burst events observed with the HRS are shown by
asterixes. It is notable that about 83 $\%$ of the burst events occur within
and about the confines
of the two bounding curves, which have the form :
\begin{eqnarray}
t_{u}=(0.5+exp[2.02-(f/62)]) \qquad s \qquad \& \nonumber \\
t_{l}=(0.05+exp[0.01-(f/57)]) \qquad s
\end{eqnarray}
Here, f is the frequency in MHz, and
$\mathrm{t_{u}~\&~t_{l}}$ are the durations
as defined by the upper and lower curves
of Fig. \ref{6fg5}, over the frequency range.\\
\\
Based on their time profile, noise storm bursts are defined as follows :
\begin{itemize}
\item{} Structures with time constants of a few hundred milliseconds.
\item{} Identifiably short-lived, narrow-band single bursts.
\item{} Typical duration, bandwidth/duration
ratio and distribution in the time-frequency plane.
\item{} Near irregular distribution in the frequency-time plane, showing
no strict time or frequency correlation among the different bursts.
\end{itemize}
The typical time profile for bursts shows
a steep rise, a short saturation phase, followed by an
exponential decay.
The essential criteria are a smooth increase and decrease in the profile.
Considering the decay phase of individual bursts,
an exponential decay law suggests a physical description in terms of a
suitable damping mechanism, either
inherent to the ambient plasma or to the exciter mechanism itself
that accounts for it.
If it can be shown that the e-fold decay times are somehow
related to the burst duration or to the peak flux values on a fixed
frequency, a relation of the damping mechanism to the exciter mechanism could
be envisaged. If, on the other hand, the decay rates do not
correlate with the burst durations, but tend to a constant value, a damping
mechanism independent of the radiation process ought to be assumed.
The time profiles of bursts exhibit a characteristic exponential decay, with
frequency dependent decay rates ( to be explained later in the section ).
This favours a damping mechanism inherent to the ambient
plasma and hence plasma emission hypothesis for the bursts.
The decay mechanism may be due to one of collisional damping,
Landau damping or deflection-time for streaming electrons.\\
\begin{figure*}
[a]{\includegraphics[totalheight=2.39in]{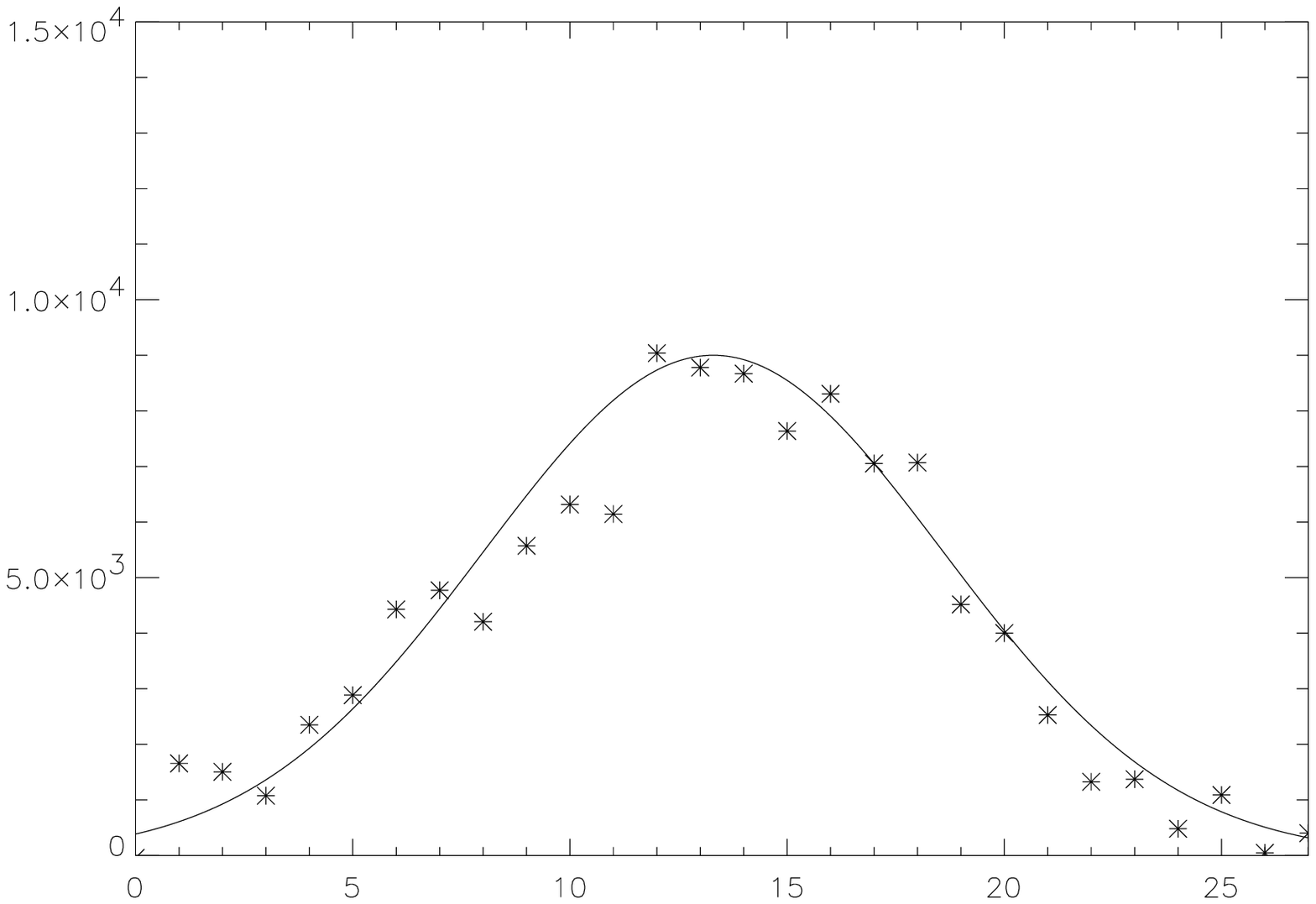}}
[b]{\includegraphics[totalheight=2.39in]{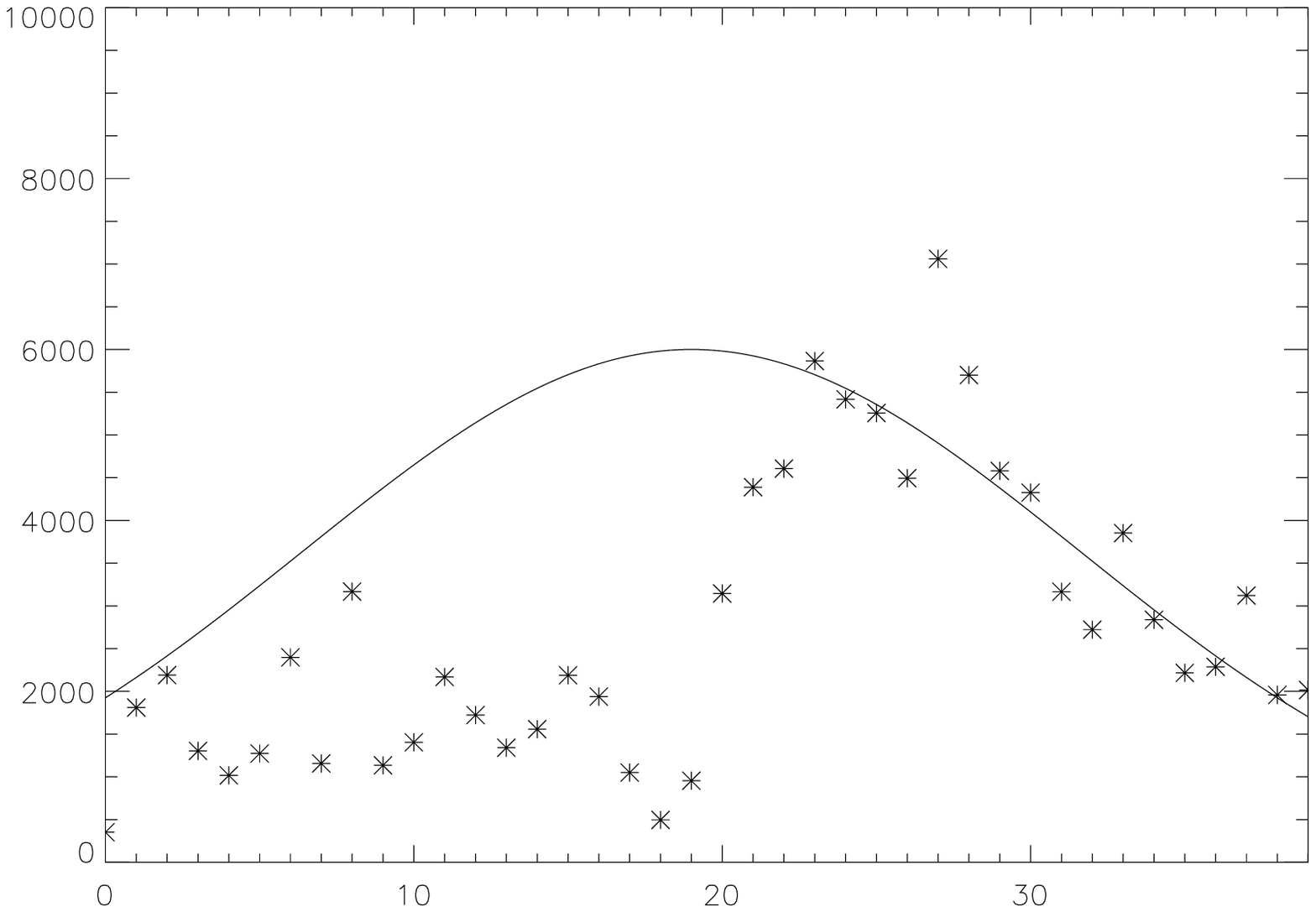}}
[c]{\includegraphics[totalheight=2.39in]{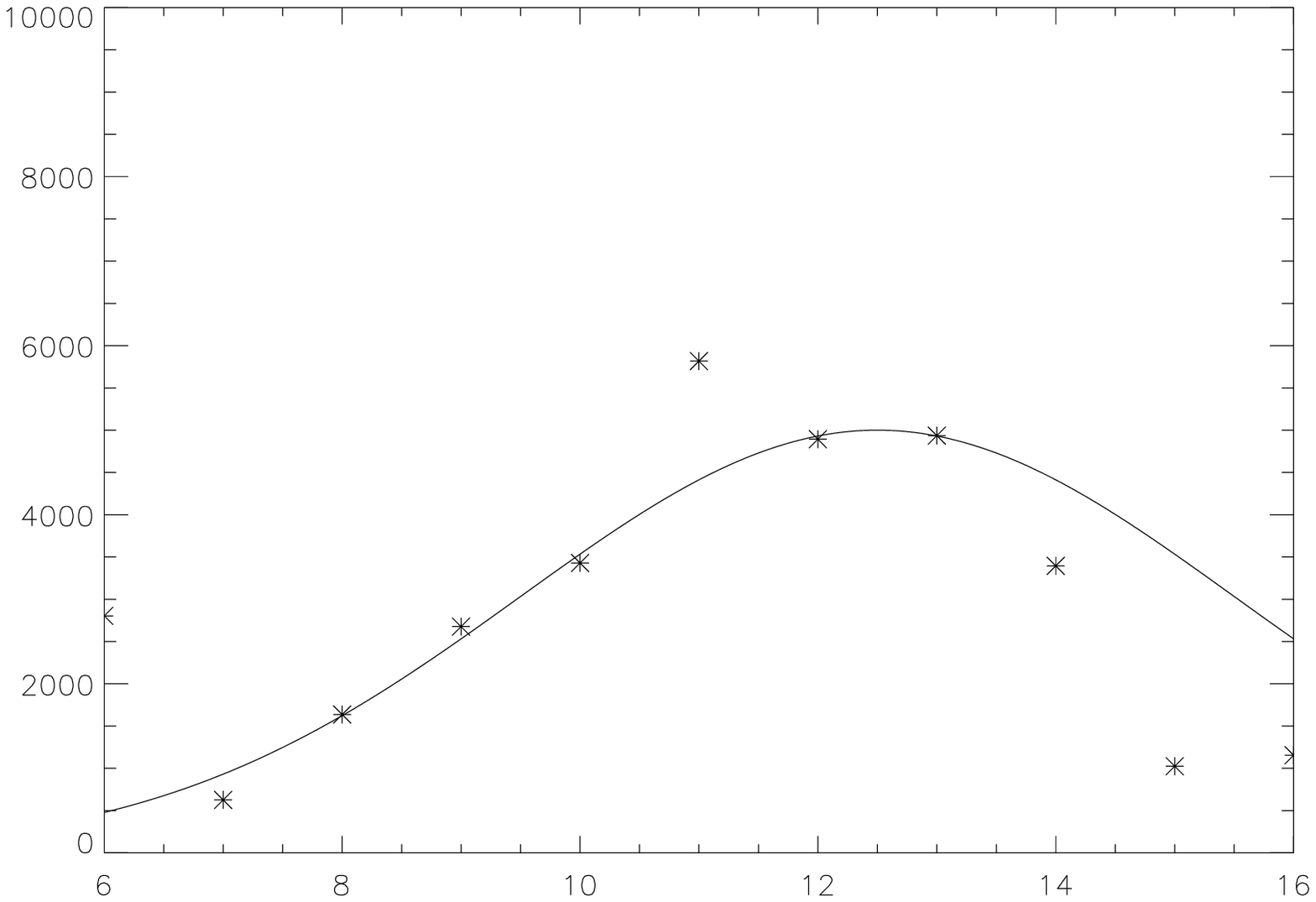}}
[d]{\includegraphics[totalheight=2.39in]{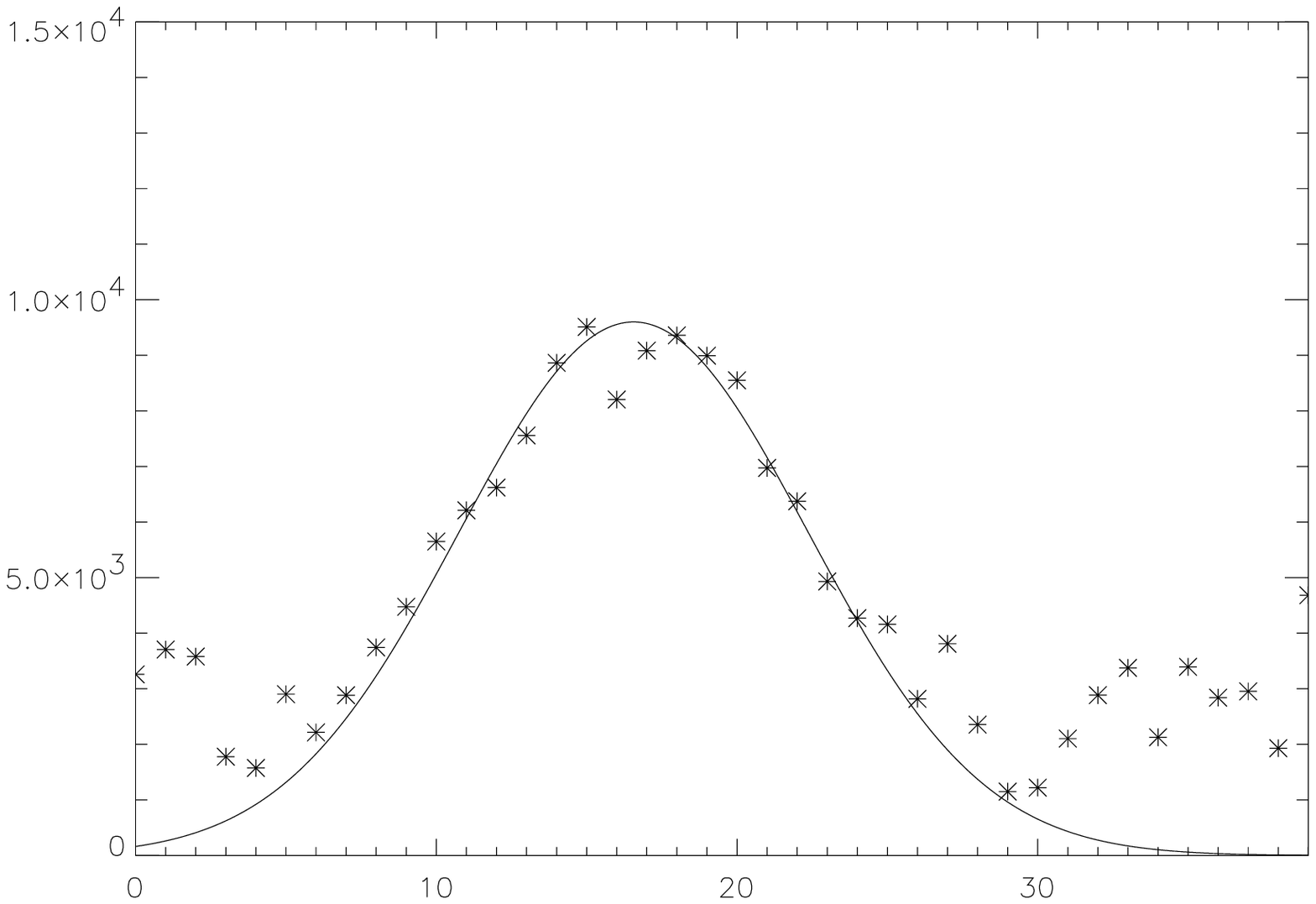}}
\caption{The 3 kinds of isolated type I noise storm bursts,
according to their averaged time profiles.
The x-axis is in units of time, with the adjacent data points
separated by $\approx$~0.25 s ; the y-axis is in
uncalibrated flux units.
Sub-plots (a) \&~(d) indicate symmetry, while
(b) \&~(c) show trailing \& leading edge
assymetry, to the fitted gaussian distribution.\label{6fg9}}
\end{figure*}
\\
\noindent
The time profile analysis involves determining the rise \& decay times
of the individual noise storm bursts.
The rise-phase does not display an exponential growth; a better description
would be that of a Gaussian profile, which
usually fits the general behaviour of the burst profiles,
often down to flux levels beyond which the decay
turns into an exponential phase.
Standard gaussians,
of the form shown in Eq. \ref{6eq3}, were fitted to the time profile
of bursts, each at their central frequency of emission. When
classified based on the degree of deviation from these gaussian profile,
the study reveals that,
$\approx 38\%$~of events had a suitable ( symmetric ) gaussian
threading the data points, and $\approx 62\%$
with steep cut-offs along the leading
( decay-phase ) or the trailing ( rise-phase ) edges of the distribution.
The plots in Fig. \ref{6fg9} explain the method invoked for time profile
studies of type I bursts. While sub-plots (a) \&~(d) trace a gaussian
distribution, sub-plots (b) \& (c) have steeper distributions of their data points
on the ascending and descending sections respectively.\\
\begin{figure}
\resizebox{\hsize}{!}{\includegraphics{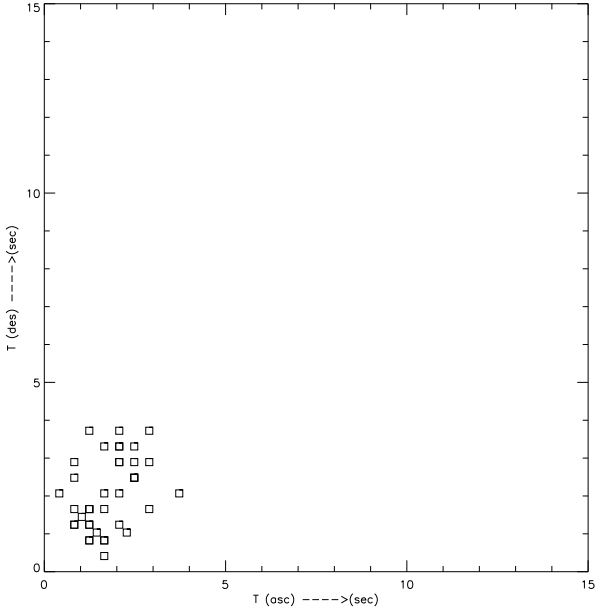}}
\caption{Scatter-plot of ascending vs. descending times of the isolated
type I noise storm bursts.\label{6fg10}}
\end{figure}
\begin{figure}
\resizebox{\hsize}{!}{\includegraphics{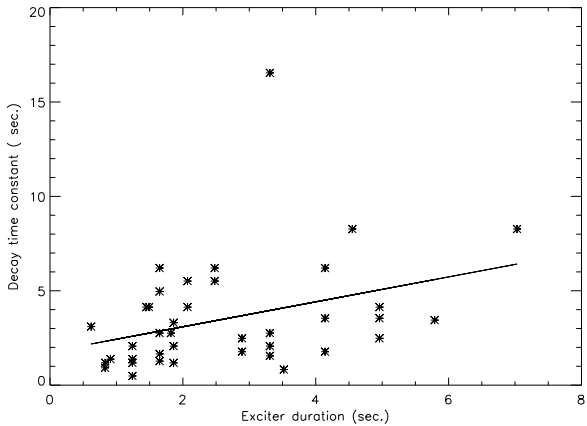}}
\caption{Scatter-plot of exciter duration vs. Decay time constant for
the isolated type I noise storm bursts.\label{6fg11}}
\end{figure}
\\
\noindent
The rise-time ($t_{r}$)~and the decay-time ($t_{d}$)~for each of the
profiles, at their central frequency of emission, was determined
based on the symmetric gaussian profile fitted to each of the
isolated noise storm burst events chosen.
Using adequate parameters, the fittings give an excellent
approximation to the real burst, ignoring the noise fluctuations.
A smoothed minimum envelope was calculated and assigned as the mean
local background to the obsevations. The background was then subtracted,
taking care not to reduce the modulated flux of the individual bursts.
Hence the times would now
correspond to the extents of the half-power points in the actual data,
from the symmetry-axes
of the fitted gaussian profiles. $t_{r}$ \& $t_{d}$~may be considered
as the characteristic time interval for the growth of plasma
instabilities (\citet{mes03, krs81} and references therein), and
the damping of the plasma waves in the source region for the bursts, respectively.
A plot of ($t_{r}$ vs. $t_{d}$), as shown in Fig. \ref{6fg10}, has a
wider dispersion of the points about the mean.
The result shows that
a majority of the burst events had either a longer phase of instability
and got quickly damped, or vice versa.
The scatter in the statistical value for the decay time,
derived from all the bursts considered, may be due to instabilities
during the excitation or to peculiarities during the decay phase.\\
\\
Since a Gaussian curve extends to infinity on the decay portion
of the time profile ( tail of the distribution ),
a fitting using a true exponential-damping curve of the form :
\begin{eqnarray}
exp[(t-t_{0})/\tau]
\end{eqnarray}
where $\tau$~is the decay time-constant and $t_{0}$~is a fitting-parameter,
was attempted, which would offer a better description of the segment
that occurs at the lower intensity level below the peak.
A semi-log plot of the intensity ( or calibrated flux S )
versus time of the
e-fold decay segment would then yield a linear result, whose slope
determines the decay ( or damping ) constant $\tau~( = - [d~ln~S/dt]^{-1})$.\\
\\
The burst duration can be attributed
to the lifetime of the coherent electron beam that excites the plasma
oscillations~\citep{tak63}. This lifetime is defined by the disruption in the orderly
motion in the electron beam, caused by their collisions with, and deflection by, the
thermal ions. The resultant energy loss by the electron beam excites coherent plasma
wave oscillations. Hence, the scale of the burst duration is comparable to this
deflection time scale, which in turn directly varies as the
third power of the electron beam velocity, implying that
the velocity of the exciting disturbance is quite small; conversely, the bursts
themselves were generated in a region of complex, filamentary coronal magnetic field
topology, where the thermal energies are relatively lower than that of the ambient
coronal plasma. For bursts of shorter duration, the value of the electron beam
velocity approaches the regime wherein the collisionless Landau damping becomes
more pronounced.\\
\\
In order to determine the actual damping phenomenon ( collisional or
Landau damping ), an estimation of the electron-ion collision temperature
$T_{c}$~was performed, based on the expression~\citep{aub72, gab90}:
\begin{eqnarray}
T_{c}~~\propto~~\nu^{4/3} \cdot \tau^{2/3}~~~~~K
\end{eqnarray}
where $\nu$~is the central frequency ( MHz ) of the isolated burst event and
$\tau$~is the damping constant in seconds. Collisional damping is sensitive
to ambient plasma temperature which varies for different events.
The values for $T_{c}$
were found to vary from $1.5 \times 10^{6}$~K to~$1.3 \times 10^{7}$~K
in the 30-130 MHz range, evidently much
less than the high brightness temperatures ($> 10^{11}$~K)~associated with
noise storm bursts~\citep{thj91}.
This, combined with the predominantly symmetrical burst distribution,
as revealed from the frequency-profile studies (Fig. \ref{6fg10}),
and the failure of the collisional damping hypothesis
to account for the bursts' decay rates, lead to the reasoning that,
the decay phase ( tail of the exciter stream ) is attributed to
non-collisional damping mechanisms.\\
\\
The gaussian distribution offers an appropriate fitting to the time profiles
of the bursts upto a point ( that also serves to locate the extent of
the exciter duration~\citep{aub72},
in this case to an accuracy of $\pm~0.4$~s. )
beyond which the exponential decay law sets-in.
A scatter-plot of the exciter duration vs. decay time constant, for the
bursts considered here, is depicted in Fig. \ref{6fg11},
along with a linear least-squares fitting.
A positive correlation of $\sim~35 \%$~is found. This
weak dependence of $\tau$~on the exciter duration, and the
fact that an exponential decay law generally favours collisional
damping of the plasma waves~\citep{aub72},
along with a value of $\Delta r$~that falls short of those obtained
from multifrequency imaging studies of type I bursts~\citep{thj91}
( implying that collision results in damping of the L-waves at a
much faster rate than the observed decay of the bursts ), could
be indicative of the collisionless Landau damping phenomenon of L-waves
by the background plasma, especially at these metric wavelengths
~\citep{zai72, elg61, mal62},
not accounting for the
case-sensitive nature of the very emission process of type I bursts.\\
\\
Applying the standard expressions for group-velocity ($v_{g}$)~and
Landau damping constant ($\gamma_{L}~(\equiv~\tau^{-1})$)~,
\citep{sga04, thj91, tha90}, and assuming values
for the phase velocity ($v_{ph}~(\approx~v_{T_{b}}))$~ to be
$10^{10}~cm~s^{-1}$~and $v_{T}~\approx~3.89~.~10^{8}~cm~s^{-1}$,
the estimated damping-length ($\Delta r$)
is $\sim~1.1 \times 10^{-2}~~R_{\odot}$.
$v_{ph}~\approx~v_{T_{b}}$,~when critical fluctuations
in $T_{eff}$~
of the L-waves,
at the threshold densities of the trapped electrons that lie proximal
to the onset of plasma instabilities, rise steeply to the levels
of $T_{b}$~observed for the type I burst events.
\section{Discussion and Conclusion}
Existence of large,
widely separated sunspot groups or active region ( AR ) complexes,
with a high degree of complexity and strength in the associated
magnetic topology,
a large-scale reorganization of magnetic
field structure in the photosphere and corona, of opposing polarity,
interconnected by dense, expansive coronal magnetic arches,
constitute the choicest of spatio-temporal correlations, for
fundamental plasma frequency radiation associated with
noise storms~\citep{mer84}.
The source region for the noise storms radiation is positioned
along the magnetic loop that appears across bipolar
ARs, and proximal to the apex of
the loop, for the magnetic field in the region's leading spot \citep{stw85}.
Imaging and spectral observations of radio noise storms strongly suggest
that there is a close relationship between bursts and energy release events
at their source~\citep{mat97}.
From the observed fine spectro-temporal characteristics of isolated
type I bursts, it is likely that they
are a signature of scattered small-scale
sites of energy release giving rise to electron acceleration
\citep{bewen81,spbeh81}.
The sites are related to the locations of
trapped electrons within the loop,
and appear as bright regions in the soft X-ray and extreme ultraviolet
images, due to the
acceleration of these nonthermal electrons to energies of a few keV to
few tens of keV. This
acts as a potential storehouse for persistent energy conversion and
acceleration of the
nonthermal electrons, contributing to the noise storms bursts at
metric wavelengths. The discussions that follow have been vividly set
to explain the varied causative mechanisms, any, or a few, or all
of which could have a definite hand in the highly transient nature of 
noise storm bursts.\\
\\
Type I noise storm bursts are structured in time, much akin to
the decimetric spikes~\citep{bak01,ksj96};
their morphological similarities
having been well-discussed in studies on the latter
\citep[see][]{kbab95,lar94,gab90,bzt88,zai72}.
One plausible
reason for the fragmented appearance could be their largely
stochastic origin.
The effective temperature (${T_{eff}}$)
of the L-waves equates
the observed brightness temperature
(${T_{b}}$)~of the burst radiation, when the
nonthermal electron beam density approaches a
threshold density value.
The values of ${T_{b}}$,
as estimated from the dynamic spectral data, are
most likely produced by trapped electrons, when the ratio of
number density of the superthermal electron beam to that
of the ambient coronal plasma
(${n_{b}/n_{e}}$) exceeds a critical threshold~\citep{tha90}.
In a trap, the presence of a loss cone
would lead to the electron distribution turning anisotropic.
At the threshold densities,
(${T_{eff}}$) of the L-waves rises steeply
\citep{the91,thj91}, and
the anisotropic distribution of these
energetic particles generate bursts at random.
The effect is identical to the critical fluctuation in
${T_{eff}}$~of the L-waves near the onset of a
plasma instability. This threshold condition is
governed by the
stochastic nature of electron acceleration whereby, they are
injected into the
source region by collisionless shock waves, that move almost at the
Alfv\'{e}n velocity \citep{bzt88,thj87}; hence the
phenomenon of noise storm bursts exhibits sporadicity.\\
\\
The nature of the fragmentation, of the kind observed as isolated
type 1 bursts,
could be due to stochastic boundary or initial conditions of
the respective processes, such as inhomogeneities in the coronal plasma,
or a low-dimensional, non-linear deterministic process capable of
generating such complicated patterns.
Single bursts can be isolated, separated by quiet phases.
The narrowband bursts show a tendency to be interrupted
in their stationary phases, due to
short-term suppressions of the emission mechanism.
The variability is caused by a stochastic input
or a high-dimensional mechanism.
In studies performed on the measured time-series of spikes,
no low-dimensional behaviour was noticed~\citep{iab94}, implying
their origin in complex processes with dimensions ( degrees of freedom )
from 4 to 6, which includes infinity ( stochasticity ),
since high-dimensional deterministic behaviour is hard to distinguish from
stochasticity.\\
\\
Suppression of the emission mechanism, for
durations comparable to that observed in sporadic burst,
could be explained as a consequence of either an inhomogeneous
plasma flowing into a reconnection region, or activity at a shock
front modulated by the upstream medium, creating cascading
magnetohydrodynamic (MHD)
turbulence in the plasma reconnection outflows \citep{lar94,sat82}.
The influence of the MHD turbulence on the resulting radio emission is
two-fold : (i) it chaotically changes the radio emission frequency, and
(ii) the very fast plasma parameter changes in the radio
source reduce or even stop the plasma instability under study, thereby
effectively interrupting the radio emission \citep{bak01,ksj96}.\\
\\
The emergence of new magnetic flux at sites of active regions causes an
increase in the size of the coronal loops that connect regions of
unlike polarity.
The newly emerging magnetic flux, and its
reconnection with the preexisting flux lines,
results in localized mass flow~\citep{mcl81}; the
ensuing weak, super-Alfv\'{e}nic shock causes wake microturbulence
\citep{spbeh81,mel80},~leading to
the excitation of plasma waves, within a few hours of
their emergence at the ARs.
The transformation of these high frequency Langmuir (L) waves into
transverse electromagnetic ( TEM ) waves occurs
as a result of their scattering and coalescence, either
on ion density fluctuations
above the sites of loop-reconnections and along the
neutral-line current sheet or with the lower-hybrid waves
excited by the trapped super-thermal ions with a loss cone, that are
borne out of weak collisionless shocks, generated due to flux
emergence~\citep{wen85,mel80}.
The reconnection loops need to be sufficiently dense and reach the
level of the local plasma
frequency, for the ambient coronal plasma to behave as a source
region for the escape of type I bursts.\\
\\
Observations of bursts of type I noise storms, at high frequency and
time resolution, have conclusively shown their origin
as a consequence of the nonthermal plasma emission mechanism.
Analyses of the frequency and time profiles of bursts,
which are a product of non-linear plasma instabilities in the source region
and governed by laws of stochastics, reveal significant insight to the
sites of large scale magnetic
reorganization and reconnection in the outer solar corona.
A drift-free nature for the narrowband emission is a sign of
high-degree homogeneity in the source region of bursts, while the
largely asymmetric time profile distribution reflects a tussle
between growth and decay of plasma wave instabilities in the turbulent
outer corona.\\
\\
The distribution of bursts with central-frequency,
bandwidth, duration and peak
radio flux, indicate that isolated type I noise
storm bursts are
intense, short duration and narrow bandwidth events,
generated at source regions
that have a low degree of non-uniformity - regions that have
a smoothly varying
density, temperature and magnetic field structure.
Studies on the narrow bandwidth of the isolated bursts conclude that,
the dimension of the exciting disturbance is of lesser extent
along the direction of the
magnetic field lines, implying a smaller velocity dispersion
and a near constant
density and temperature structure. This reveals the
high degree of uniformity
in the source region for the noise storm bursts;
and since the corona above the active
regions is in a high state of inhomogeneity, the source
region for bursts needs to be
of a relatively smaller spatial extent.\\
\\
Possible alternative explanations, for the
weak dependence of $\tau$~on the exciter duration
would be cases wherein, for the same exciter length,
their velocities are larger, they encountered
large density gradients along their trajectory ( small duration for
the exciter, or lower temperatures ) - verifiable
from white-light coronograph observations, or the turbulent
bandwidth ( exciter duration ) is larger at increased
coronal temperatures.
From the physical emission process point of view :
\begin{itemize}
\item{} The exciter has a finite length as it travels through
the corona. The time profile then would reflect the crossing
of the exciter through the layer corresponding to the observed channel
bandwidth ( assuming a plasma hypothesis ).
\item{} The exciting agent emits
simultaneously on all frequencies where the burst is observable
corresponding to a plasma layer of given height. The exciter function then would
represent purely a temporal description of the emission process.
\item{} The time profile may be a combination of both: if a travelling disturbance
starts a characteristic emission process going on even after the crossing
of the agent through the observed layer, the resulting time profile
could be a convolution of the time
profile of the stimulation-agent and the induced emission.
\end{itemize}
\noindent
The total bandwidth of observed bursts is found to be 2-5 \% of the
central frequency; hence, the longitudinal extent of the emission region
of a burst is several times less than the total beam length,
assuming radiation to be near the local plasma frequency.
This suggests that the beam emits only in a narrow
region. After traversing it, the beam may either have decayed to a stable
distribution, or the medium is unable to produce radio waves.
Since the burst emission is triggered by a beam of superthermal
electrons propagating through the corona ( with a spread in their
velocities ), whose density exceeds a
certain threshold leading to plasma instabilities at the source region,
the reason for ceasation of bursts could be due to stabilization in
the causative electron beam as a result of non-uniformity in
velocity ( Landau damping of plasma waves due to an excess of electrons
with velocities that are smaller than the phase velocity of the wave )
and density profile~\citep{zai72}, which in themselves are strongly
governed by stochastic considerations.\\
\\
The details on type I emission process are expected to critically depend
on (i) local physical conditions in the emitting sources, which are likely
to vary with the ARs, and even within the magnetic field structure
associated
with one given AR (ii) the spread in intensity of the bursts'
emission itself.
Thus, if the distribution of their peak flux densities is strongly affected
by the emission process, it would change significantly with the storm event
and frequency. Such observable characteristics lead to a fair speculation as
to the peak flux density of each burst being nearly linearly related to
the energy transferred to the nonthermal electrons.
Characteristics like stationarity, intermittency and dimensionality
need to be taken into account, in order to interpret the
free magnetic energy
release mechanism in the solar corona due to the noise storm bursts.
Added
information regarding the plasma structure in the emission region and the
spectral behaviour of the bursts would be an essential pre-requisite for
such a study.
\section*{Acknowledgments}
We wish to thank E. Ebenezer of the
Indian Institute of Astrophysics for system implementation and data
archival in the context of the HRS.
We also thank the referee for the effort and insightful comments
that have improved the clarity in content and presentation of this paper.
We are immensely grateful to the scientific and technical staff
of the Gauribidanur Radio Observatory, jointly run by the
Raman Research Institute and the Indian Institute of Astrophysics,
for duly maintaining the scientific facilities on-site.
\appendix
\section[]{Calibration of Dynamic Spectrum using the Galactic background}
The method employed in calibration of the flux densities of
type I bursts, over the
entire frequency range of the spectrograph, to an absolute scale,
is relevant to
low-frequency (${\le 100}$~MHz ) radio telescopes,
with inherent low angular resolution, used in spectral observations of
solar radio bursts, with the galactic background
radiation as the principal contributor ( in comparison to receiver noise )
to the system temperature.
The antenna response to the brightness  distribution of the
galactic background
radiation is a function of its angular resolution and observing frequency.
The spectrum
of this unpolarized galactic background has been well catalogued
\citep{can79, nov78, brn73, alx69, ytw66},
in either of the galactic hemispherical regions, and
serves as a highly reliable calibrator source at metric
\& decametric (1-100 MHz) wavelengths. Since radio signals from the
Sun and the
galactic-plane essentially arrive along the same path, in their
approach to the
receiving system, this technique ensures that the errors and
losses introduced along the
signal path ( like impedance or ohmic-losses in the antenna system )
are effectively identical in
either case. This scheme is also
adept at correcting for ionospheric absorption and partial ground reflection
of the celestial radio signals~\citep{duk01}, to an accuracy of
2 $\%$~, and is
especially suited for spatially low-resolving telescopes,
that have difficulty in discerning the fine spatial structures in
standard, unresolvable radio sources, for use in absolute
calibration of radio data.\\
\\
For antennas with low
directivity, the galactic background radiation is essentially isotropic.
The solid angle of the group-beam of the array is 0.56-0.13 sr in the
30 -130 MHz range, while it
is 1.64 sr for the individual LPDAs in the group.
The large value for beam width implies that, variability in the
galactic background
with LST, as contributed by the less intense, isotropic,
large-scale features, remain minimal, with the variations in sky temperature
being ${< 30 }$~per cent~\citep{ytw66} at high galactic latitudes.\\
\\
The spectra of the galactic background radiation, at the local
meridian-transit point, were obtained from observing regions that
were far removed from the galactic plane - ideally
towards the North Galactic Pole ( NGP ) and the
South Galactic Pole ( SGP ); thereby
large changes in the brightness distributions due to
structures ( like partially-ionized hydrogen ( HI )
in the interstellar medium )
located in the galactic plane, get excluded.
The NGP ( J2003.0~:~${\alpha=12^{h}51^{m}33.^{s}9675,
\delta=+27^{o}06^{'}49.^{"}3824}$~) was observed on
2002 December 25 and the SGP
( J2003.0~:~${\alpha=00^{h}51^{m}33.^{s}9675,
\delta=-27^{o}06^{'}49.^{"}3824}$~)
the previous day, the former at early dawn ( LST=11:21:34 ; IST=05:28:30 )
and the latter around local sunset ( LST=23:21:34 ; IST=17:30:27 ),
in the transit mode. This observation schedule augured well with the
location-coordinates and beam width of the LPDAs in the
four-antennas group, as to
pursue the observations and determine the spectrum in the direction
of the two poles.
The observing epoch was carefully chosen to ensure that
the Sun had a minimal influence in the primary beam response.
Considering the tapering-off of gain ( due to array-pattern response )
with angular distance away
from the local meridian at zenith, a symmetrical observing window
of about 3 hrs
was chosen. Since the presence of large sidelobes,
especially at the lower frequencies of the observing spectral range,
would lead to the acquisition of remnant solar flux
proximal to the horizon,
only 1 h of the galactic background
about the transit point remained utilizable in the calibration of
solar data.
The wider beam along the E-plane meant that `boresight' alignment
to declination
values for the poles alone needed to be set, so that the antenna
temperature is
contributed predominantly by the brightness distribution of the
galactic background. Results from observations of the galactic
background during the course of the day, at either of the poles,
were found to have good agreement with similar
data published from observations done previously.\\
\\
The availabilty of 1 h of data on the calibration source,
symmetrical about the
transit point, set an identical limit on utilization of the solar
spectral data, for the specific days of observation.
The gain across all the 401 frequency channels was found to have
reasonable stability.
With a knowledge of the deviation of the observed spectrum
from the catalogued spectra of the galactic background at
identical LSTs, and
the antenna gain in the direction of the radio source,
the antenna temperature was converted to the noise storm
flux density values.
For the total-power receiver case,
the detected signal is a convolved response of the antenna
power pattern and the
sky brightness distribution, across a channel width of 250 kHz,
termed the
spectral power, measured in units of ${mV^{2}/Hz}$.
The spectral power is then transformed to units of
radio flux density ( measured in W ${m^{-2} Hz^{-1}}$ ),
given the data on the spectral
power of noise storms and the galactic background,
and the model for radio flux density of the latter.

\bsp

\label{lastpage}

\end{document}